\documentclass[prb,twocolumn,showpacs]{revtex4}
\usepackage{graphicx}
\usepackage{amsmath}
\usepackage{amssymb}

\def\lsim{\
  \lower-1.2pt\vbox{\hbox{\rlap{$<$}\lower5pt\vbox{\hbox{$\sim$}}}}\ }
\def\gsim{\
  \lower-1.2pt\vbox{\hbox{\rlap{$>$}\lower5pt\vbox{\hbox{$\sim$}}}}\ }

\begin{document}

\title{Calculation of the He-II quasiparticle spectrum by the method of collective
variables}

\author{M.D.~Tomchenko}
\affiliation{Bogolyubov Institute for Theoretical Physics\\
        14b, Metrologichna Str., Kyiv 03143, Ukraine\\ e-mail: mtomchenko@bitp.kiev.ua}

\date{\today}

\begin{abstract}
The method of collective variables (MCV) has been used to
calculate the logarithm of the He-II ground-state wave function,
$\ln{\Psi_{0}}$, to an accuracy of a first correction to the
Jastrow function and, in a second approximation, the wave function
$\Psi_{\mathbf{k}}$ of the first excited
state and the He-II quasiparticle spectrum. The functions $\Psi_{0}$ and $%
\Psi_{\mathbf{k}}$ were found as the eigenfunctions of the $N$-particle Schr%
\"{o}dinger equation, and the function $\Psi_{0}$ was connected to
the structure factor of He-II, using the Vakarchuk equation. The
model does not contain any fitting parameter or function. The
quasiparticle spectrum calculated numerically agrees well with the
experiment. Our solution improves the result obtained early by
Yukhnovskyi and Vakarchuk.
\end{abstract}



\maketitle

\section{Introduction}

The structures of the $N$-particle wave functions of the ground and weakly
excited states of helium-II are known in the main \cite%
{1,2,3,4,5,6,7,8,9,10,11}, and the solutions which take into account several
first approximations have been obtained. In our opinion, the main unresolved
problems concerning the microscopic physics of He-II are the structure of
the composed condensate, the nature of the $\lambda$-transition, and the
role of microscopic vortex rings.

The form of the He-II quasiparticle spectrum has been forecasted by Landau
for the first time \cite{13}. In Feynman's known works \cite{1,2,3}, an
opportunity to determine this spectrum making use of the structure factor
has been demonstrated. Feynman intuitively found the structure of the $\Psi_{%
\mathbf{k}}$ wave function for the state of He-II with a single phonon and
approximately deduced the He-II quasiparticle spectrum. According to Feynman
and Cohen,
\begin{equation}
\Psi_{\mathbf{k}}(\mathbf{r}_{1},\ldots,\mathbf{r}_{N})=\psi_{\mathbf{k}}(%
\mathbf{r}_{1},\ldots,\mathbf{r}_{N})\Psi_{0}(\mathbf{r}_{1},\ldots ,\mathbf{%
r}_{N}),   \label{1}
\end{equation}%
\begin{equation}
\psi_{\mathbf{k}}=\rho_{-\mathbf{k}}+\sum\limits_{\mathbf{k}_{1}}^{\mathbf{k}%
_{1}\neq0,\mathbf{k}}A\frac{\mathbf{k}_{1}\mathbf{k}}{k_{1}^{2}}\rho_{%
\mathbf{k}_{1}-\mathbf{k}}\rho_{-\mathbf{k}_{1}},   \label{2}
\end{equation}
where $\Psi_{0}$ is the wave function of the ground state,
\begin{equation}
\rho_{\mathbf{k}}=\frac{1}{\sqrt{N}}\sum\limits_{j=1}^{N}e^{-i\mathbf{k}%
\mathbf{r}_{j}}\quad(\mathbf{k}\not =0)   \label{rok}
\end{equation}
are collective variables \cite{14}, and $N$ is the total number of atoms in
helium. However, it has not been shown in Feynman's works that function (\ref%
{2}) is the eigenfunction of the $N$-particle Schr\"{o}dinger equation.
Feynman's ideas have been developed in a great number of works (see, e.g.,
\cite{15,16,17,18,19,20,21}). The Feynman-Cohen function has been specified
in works \cite{5} where the analysis of the total Hamiltonian of the system
has been carried out. A more accurate form of the function $\psi_{\mathbf{k}}
$ and the structure of the function $\Psi_{0}$ have been found in works \cite%
{6,7,8,9,10,11}, where $\Psi_{0}$ and $\psi_{\mathbf{k}}$ were sought as the
eigenfunctions of the Schr\"{o}dinger equation.

The idea of the MCV has been proposed in Bogolyubov and Zubarev's work \cite%
{14}. This method has been substantiated and developed in works \cite%
{7,8,9,10,22,23}. In work \cite{11}, taking advantage of the MCV, the $%
\Psi_{0}$ and $\Psi_{\mathbf{k}}$ functions of helium-II have been
calculated making use of the model potential of interaction between He$^{4}$
atoms with one fitting parameter. Nevertheless, as was indicated in \cite%
{9,10,11}, the derivation of the $\Psi_{0}$ and $\Psi_{\mathbf{k}}$ wave
functions, as well as the He-II quasiparticle spectrum, starting from the
He-II structure factor, known from the experiment, rather than from the
model potential has significant advantages. In this case, the problem does
not contain fitting parameters, and one can avoid the task of description of
atomic interaction at small distances, which arises because of atoms'
extension \cite{9,10,11}.

Such an approach has been considered in work \cite{8}, where $\ln{\Psi_{0}}$
was found in a zeroth-order approximation, while $\psi_{\mathbf{k}}$ and the
He-II quasiparticle spectrum in a first one. The obtained spectrum $E(k)$
agreed well with the experiment. In this work, we calculated $\Psi_{0}$, $%
\psi_{\mathbf{k}}$, and $E(k)$ more accurately. Namely, we found a first
correction to $\ln{\Psi_{0}}$ and a second ones to $\psi_{\mathbf{k}}$ and $%
E(k)$. In doing so, we used the equation for $\Psi_{0}$, derived in work
\cite{22} (below, we coin it as the Vakarchuk equation). Actually, the
expansion parameter of the problem was the function $2\sigma(k)k/k_{0}$ (see
Fig.~1), the average value of which within the interval $k=0\div k_{0}$ was
about $-1/2$; i.e. the parameter was not small. Therefore, the corrections
to $\ln{\Psi_{0}}$ and $E(k)$, generally speaking, were not small too, and
their calculation was of interest.

\section{The ground state of helium-II}

A more detailed analysis of the equations and the method of determining $%
\Psi_{0}$ were exposed in works \cite{9,10,11}. The necessary equations for $%
\Psi_{0}$ and $\psi_{\mathbf{k}}$ were found by Yukhnovskyi and Vakarchuk
\cite{8,9,10,22}. We shall use different notations and different forms of
the equations for $\Psi_{0}$ and $\psi_{\mathbf{k}}$ \cite{11} (the latter
is partially caused by our desire to reduce the error of numerical solution
of the equations \cite{11}).

The wave function of the ground state of He-II is sought in the form \cite%
{9,11}
\begin{equation}
\Psi_{0}=e^{{\small S_{0}}},   \label{4}
\end{equation}%
\begin{equation}
S_{0}=\sum\limits_{\mathbf{k}\not =0}\sigma(k)\rho_{\mathbf{k}}\rho _{-%
\mathbf{k}}+\sum\limits_{\mathbf{k}_{1},\mathbf{k}_{2}\neq0}^{\mathbf{k}_{1}+%
\mathbf{k}_{2}\not =0}\frac{f(\mathbf{k}_{1},\mathbf{k}_{2})}{\sqrt{N}}\rho_{%
\mathbf{k}_{1}+\mathbf{k}_{2}}\rho_{-\mathbf{k}_{1}}\rho_{-\mathbf{k}_{2}}.
\label{5}
\end{equation}
The corrections of higher orders to $S_{0}$ [Eq.~(\ref{5})] are neglected.
In this approximation, the relation
\begin{equation}
f(\mathbf{k}_{1},\mathbf{k}_{2})=-\frac{2\sigma(k_{1})2\sigma(k_{2})\mathbf{k%
}_{1}\mathbf{k}_{2}}{e(\mathbf{k}_{1}+\mathbf{k}_{2})+e(k_{1})+e(k_{2})},
\label{6}
\end{equation}
where
\begin{equation}
e(\mathbf{k})=k^{2}(1-4\sigma(k)),   \label{7}
\end{equation}
is valid \cite{11}. In works \cite{22}, an equation that connects $\Psi_{0}$
[Eqs.~(\ref{4}) and (\ref{5})] with the He-II structure factor $S(k)$ was
derived. We shall write down this equation in approximation (\ref{5}) for $%
\Psi_{0}$ and using the notations of work \cite{11} as follows:
\begin{equation}
4\sigma(q)=1-\frac{1}{S(q)}-\Sigma(q),   \label{8}
\end{equation}%
\begin{equation}
\Sigma(q)=\frac{1}{N}\sum\limits_{\mathbf{k}\not =0}\frac{8\sigma (k)\sigma(%
\mathbf{k}+\mathbf{q})+R(\mathbf{k},\mathbf{q})}{[1-4\sigma (k)][1-4\sigma(%
\mathbf{k}+\mathbf{q})]},   \label{9}
\end{equation}
where
\begin{equation}
R(\mathbf{k},\mathbf{q})=4f_{s}(\mathbf{k},\mathbf{q})\left[ 1+2f_{s}(%
\mathbf{k},\mathbf{q})\right] ,   \label{10}
\end{equation}%
\begin{equation}
f_{s}(\mathbf{k},\mathbf{q})=f(\mathbf{k},\mathbf{q})+f(-\mathbf{k}-\mathbf{q%
},\mathbf{k})+f(-\mathbf{k}-\mathbf{q},\mathbf{q}).   \label{11}
\end{equation}
Equations (\ref{8}) and (\ref{9}) were derived in work \cite{22} from the
known equation, which connects $S(k)$ with the pair distribution function $%
F_{2}(r)$:
\begin{equation}
S(k)=1+n\int\left( F_{2}(r)-1\right) e^{-i\mathbf{k}\mathbf{r}}d\mathbf{r},
\label{12}
\end{equation}
where $n$ is the concentration of helium atoms.

\begin{figure}[h]
\centerline{\includegraphics[width=85mm]{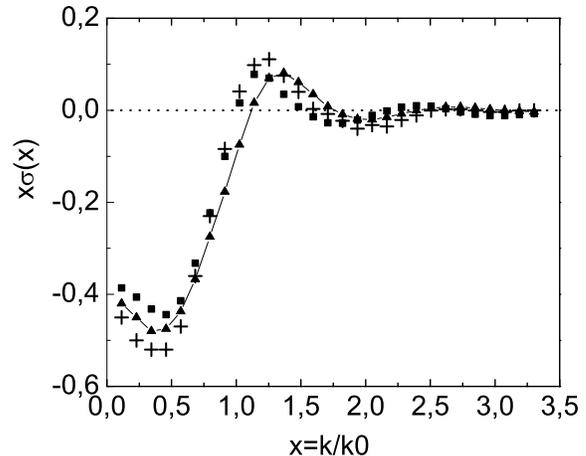}} \caption{Dependences of the quantity $k\sigma$ on $k$, $k$ being in terms of $%
k_{0}=2\pi/d=1.756~\mathrm{\mathring{A}}^{-1}$ units and $d$ the
average distance between He-II atoms. Squares mark a zeroth-order
approximation [Eq.~(\ref{14})] for $\sigma(k),$ pluses a first
approximation, and triangles a second approximation calculated
from the model \textquotedblleft elliptic\textquotedblright\
potential with $U(0)=60~\mathrm{K}$ \cite{11}.}
\end{figure}

We note that in Eqs.~(\ref{8}) and (\ref{9}), the interaction between He$^{4}
$ atoms does not present explicitly, and $\Psi_{0}$ is connected with the
He-II structure factor only, so that such a way of finding $\Psi_{0}$ allows
the problem of descriptions of interaction between He$^{4}$ atoms at small
distances \cite{9,11} to be avoided partially: provided strong overlapping
of He$^{4}$ atoms, the description of atomic interaction using the
interaction potential becomes inaccurate, because the atomic structure
becomes important under such conditions, and it is necessary, generally
speaking, to solve a quantum-mechanical problem of interaction of two nuclei
and four electrons. The function $\Psi_{0}$ [Eqs.~(\ref{4}) and (\ref{5})],
found from Eqs.~(\ref{6})--(\ref{11}), takes short-range correlations into
account more correctly than that found from the model potential \cite{11}.
For a quite correct account of the atomic structure, one should determine
the function $\Psi_{0}$ for a system of nuclei and electrons rather than $N$
structureless particles. It is a hopeless task. Nevertheless, as one can see
below, configurations with atom overlapping are very improbable, so that
from the physical point of view, it is quite reasonable to consider atoms as
structureless particles.

A single shortcoming made in the course of derivation of $\Psi_{0}$ from
Eqs.~(\ref{4})--(\ref{11}) was the break of series (\ref{5}). But, since the
model does not contain fitting parameters, the accuracy of approximation (%
\ref{5}) can be estimated by comparing both the theoretical spectrum of
He-II quasiparticles and the theoretical potential of interaction between He$%
^{4}$ atoms with experimental ones.

In order to find the wave function of the ground state, one has to know $S(k)
$ at the temperature $T=0~\mathrm{K}$. As far as we know, the most exact
measurements of $S(k)$ were carried out in work \cite{24}. We used the
smoothed data on $S(k),$ obtained at $T=1\mathrm{~K}$ in \cite{24}, and
calculated the dependences $S(k,T=0)$ by the formula \cite{25}
\begin{equation}
S(k,T=0)=S(k,T)\tanh{\frac{E(k)}{2k_{B}T}}.   \label{13}
\end{equation}
At $k\leq0.2~\mathrm{\mathring{A}}^{-1}$, we supposed that $S(k,T=0)\sim k$
(because $S(k=0,T=0)=0$ \cite{26} and $E(k\rightarrow0)=ck$ in Eq.~(\ref{13}%
)). In works \cite{8,23}, the integral equations (\ref{8}) and (\ref{9})
were not solved and a zeroth-order approximation
\begin{equation}
4\sigma(q)=1-\frac{1}{S(q)}   \label{14}
\end{equation}
was used to determine $\Psi_{0}$.
\begin{figure}[h]
\centerline{\includegraphics[width=85mm]{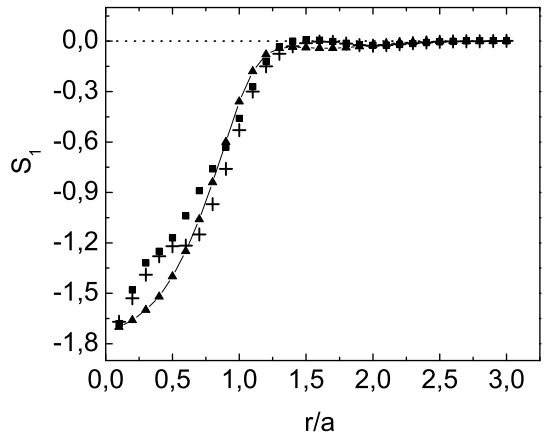}} \caption{Function $S_{1}(r/a)$ [Eq.~(\ref{15})], where $a=2.64~\mathrm{%
\mathring{A}}$ is the \textquotedblleft
diameter\textquotedblright\ \cite{33} of the He$^{4}$ atom. The
notations correspond to the same approximations for $\sigma(k)$ as
in Fig.~1.}
\end{figure}

Below, Eqs.~(\ref{6})--(\ref{11}) will be solved numerically, and the
solution $\sigma(k)$, which includes a single correction to the zeroth-order
approximation (\ref{14}), will be obtained; therefore, we shall call this
solution a first approximation to $\sigma(k)$. The solution of the integral
equation (\ref{8}) cannot be found by the iteration method, so that we used
the Newton one \cite{27} for this purpose. As a result, two solutions were
obtained, one of which, with a smaller energy per atom $E_{0}=-1.4~\mathrm{K}
$ ($E_{0}=0.1~\mathrm{K}$ in a zeroth-order approximation and $-7.16~\mathrm{%
K}$ in the experiment), being taken as the ground state. This solution for $%
\sigma(k)$ is shown in Fig.~1.

A significant body of information concerning the properties of $\Psi_{0}$ is
included into the function
\begin{equation}
S_{1}(r)=\frac{1}{N}\sum\limits_{\mathbf{k}}\sigma(k)e^{i\mathbf{k}\mathbf{r}%
},   \label{15}
\end{equation}
where
\begin{equation}
\sum\limits_{\mathbf{k}\not =0}\sigma(k)\rho_{\mathbf{k}}\rho_{-\mathbf{k}%
}=\sum\limits_{i,j}S_{1}(\mathbf{r_{i}}-\mathbf{r_{j}}).   \label{S12}
\end{equation}
The behavior of $S_{1}(r)$ at $r\rightarrow0$ shows how quickly the function
$\Psi_{0}$ decays if the atoms overlap. Fig.~2 represents the function $%
S_{1}(r)$ for a zeroth-order (\ref{14}) and a first approximation for $%
\sigma(k)$, and for $\sigma(k)$ found in a second approximation, starting
from the model potential \cite{11}. One can see that $S_{1}(0)\approx-1.7$
in all those cases. One can separate a two-particle summand of the form $%
\sum\limits_{\mathbf{k}\not =0}\tilde{\sigma}(k)\rho_{\mathbf{k}}\rho_{-%
\mathbf{k}}$ from the addend in the r.h.s. of (\ref{5}), see \cite{11}, but
a calculation shows that the account of $\tilde{\sigma}(k)$ renormalizes $%
S_{1}(r)$ very slightly, by a few percent only. Thus, provided that two He$%
^{4}$ atoms overlap, the wave function of the ground state diminishes by a
factor of $e^{3.4}\approx30$, so that a sharp reduction does not occur,
although the probability density $|\Psi_{0}|^{2}$ decreases rather strongly,
by a factor of 1000. It means that the He$^{4}$ atom possesses properties
which are intermediate between \textquotedblleft soft\textquotedblright- and
\textquotedblleft hard\textquotedblright-core ones. In case, for example,
that 10 pairs of atoms overlap, $\Psi_{0}$ decreases by a factor of $%
e^{34}\sim10^{14}$ as compared to its value for a uniform distribution of
atoms without overlapping. Therefore, configurations where many atoms
overlap are extremely improbable.

\section{Calculation of the He-II quasiparticle spectrum}

Knowing $\Psi_{0}$, one can find the wave function $\Psi_{\mathbf{k}}=\psi_{%
\mathbf{k}}\Psi_{0}$, which describes the state of the system with a single
quasiparticle of the phonon type, and the quasiparticle spectrum $E(k)$ from
the following equations \cite{11}:
\begin{align}
\psi_{\mathbf{k}} & =\rho_{-\mathbf{k}}+\sum\limits_{\mathbf{k}_{1}}^{%
\mathbf{k}_{1}\neq0,\mathbf{k}}\frac{P(\mathbf{k},\mathbf{k}_{1})}{\sqrt {N}}%
\rho_{\mathbf{k}_{1}-\mathbf{k}}\rho_{-\mathbf{k}_{1}}+  \notag \\
& +\sum\limits_{\mathbf{k}_{1},\mathbf{k}_{2}\neq0}^{\mathbf{k}_{1}+\mathbf{k%
}_{2}\not =\mathbf{k}}\frac{Q(\mathbf{k},\mathbf{k}_{1},\mathbf{k}_{2})}{N}%
\rho_{\mathbf{k}_{1}+\mathbf{k}_{2}-\mathbf{k}}\rho_{-\mathbf{k}_{1}}\rho_{-%
\mathbf{k}_{2}}+\ldots.   \label{17}
\end{align}%
\begin{align}
\tilde{E}(k) & =e(k)+\int d\mathbf{k}_{1}P(\mathbf{k},\mathbf{k}_{1})2%
\mathbf{k}_{1}(\mathbf{k}-\mathbf{k}_{1})+  \notag \\
& +\int d\mathbf{k}_{1}(-2k_{1}^{2})\left[ Q(\mathbf{k},\mathbf{k}_{1},-%
\mathbf{k}_{1})+2Q(\mathbf{k},\mathbf{k},\mathbf{k}_{1})\right] ,
\label{18}
\end{align}%
\begin{align}
& P(\mathbf{k},\mathbf{k}_{1})\left[ e(k_{1})+e(\mathbf{k}-\mathbf{k}_{1})-%
\tilde{E}(k)\right] +\int d\mathbf{k}_{2}F(\mathbf{k},\mathbf{k}_{1},\mathbf{%
k}_{2})=  \notag \\
& =4\sigma(k_{1})\mathbf{k}\mathbf{k}_{1}+2k^{2}f_{s}(\mathbf{k}_{1},\mathbf{%
k}-\mathbf{k}_{1}),   \label{19}
\end{align}%
\begin{align}
F(\mathbf{k},\mathbf{k}_{1},\mathbf{k}_{2}) & =4\mathbf{k}_{2}(\mathbf{k}-%
\mathbf{k}_{1}-\mathbf{k}_{2})Q(\mathbf{k},\mathbf{k}_{1},\mathbf{k}_{2})+
\notag \\
& +2\mathbf{k}_{2}(\mathbf{k}_{1}-\mathbf{k}_{2})Q(\mathbf{k},\mathbf{k}_{1}-%
\mathbf{k}_{2},\mathbf{k}_{2}),   \label{20}
\end{align}%
\begin{align}
& Q(\mathbf{k},\mathbf{k}_{1},\mathbf{k}_{2})\left[ e(k_{1})+e(k_{2})+e(%
\mathbf{k}-\mathbf{k}_{1}-\mathbf{k}_{2})-\tilde{E}(k)\right] =  \notag \\
& =P_{s}(\mathbf{k},\mathbf{k}_{1}+\mathbf{k}_{2})\ast G(\mathbf{k}_{1},%
\mathbf{k}_{2})+L(\mathbf{k},\mathbf{k}_{1},\mathbf{k}_{2}),   \label{21}
\end{align}%
\begin{equation}
P_{s}(\mathbf{k}_{1},\mathbf{k}_{2})=P(\mathbf{k}_{1},\mathbf{k}_{2})+P(%
\mathbf{k}_{1},\mathbf{k}_{1}-\mathbf{k}_{2}),   \label{22}
\end{equation}%
\begin{align}
G(\mathbf{k}_{1},\mathbf{k}_{2}) & =\left[ 2\sigma(k_{1})\mathbf{k}%
_{1}+2\sigma(k_{2})\mathbf{k}_{2}\right] (\mathbf{k}_{1}+\mathbf{k}_{2})+
\notag \\
& +2(\mathbf{k}_{1}+\mathbf{k}_{2})^{2}f_{s}(\mathbf{k}_{1},\mathbf{k}_{2}),
\label{23}
\end{align}%
\begin{figure}[h]
\centerline{\includegraphics[width=85mm]{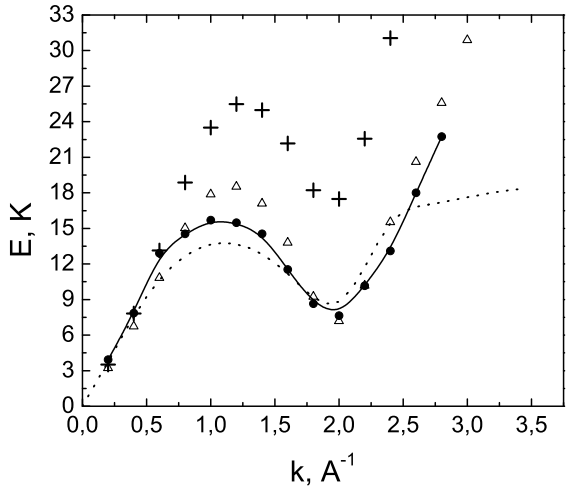}}
\caption{Theoretical He-II quasiparticle spectrum. Pluses
correspond to a
zeroth-order approximation (\ref{25}) for $\Psi_{0}$ and $\psi_{\mathbf{k}}$%
, triangles to a zeroth-order approximation for $\Psi_{0}$ and a
first one for $\psi_{\mathbf{k}}$, circles to a first
approximation for $\Psi_{0}$ and a second one for
$\psi_{\mathbf{k}}$; the solid curve is drawn using the spline
method; the dotted curve corresponds to the experimental spectrum
\cite{28}.}
\end{figure}
\begin{equation}
L(\mathbf{k},\mathbf{k}_{1},\mathbf{k}_{2})=2\mathbf{k}(\mathbf{k}_{1}+%
\mathbf{k}_{2})f_{s}(\mathbf{k}_{1},\mathbf{k}_{2}).   \label{24}
\end{equation}
In (\ref{17}), the corrections of higher orders to $\psi_{\mathbf{k}}$ are
neglected. In Eqs.~(\ref{18})--(\ref{24}, we converted to the dimensionless
variables $k^{\prime}=k/k_{0}$ and $\tilde{E}(k^{\prime})=\frac{E(k)2m}{%
\hbar^{2}k_{0}^{2}}$, where $k_{0}=2\pi/d$ and $d=3.578~\mathrm{\mathring{A}}
$ is the average interatomic distance. The primes will be omitted below.

The quasiparticle spectrum, calculated using $\Psi_{0}$ and $\psi_{\mathbf{k}%
}$ in a zeroth-order approximation (i.e. $f(\mathbf{k}_{1},\mathbf{k}_{2})=0$%
, Eq.~(\ref{14}), and $\psi_{\mathbf{k}}=\rho_{-\mathbf{k}}$), has the form
of Feynman's known formula \cite{1}, which describes the spectrum of a
slightly nonideal Bose gas \cite{14},
\begin{equation}
E(k)=\frac{\hbar^{2}k^{2}}{2mS(k)}.   \label{25}
\end{equation}
This spectrum is represented by pluses in Fig.~3.

To find $\psi_{\mathbf{k}}$ in a first approximation, we should assume that $%
Q(\mathbf{k},\mathbf{k}_{1},\mathbf{k}_{2})=0$. From Eq.~(\ref{19}), we have
\begin{equation}
P(\mathbf{k},\mathbf{k}_{1})=\frac{4\sigma(k_{1})\mathbf{k}\mathbf{k}%
_{1}+2k^{2}f_{s}(\mathbf{k}_{1},\mathbf{k}-\mathbf{k}_{1})}{e(k_{1})+e(%
\mathbf{k}-\mathbf{k}_{1})-\tilde{E}(k)}.   \label{26}
\end{equation}
The system of equations (\ref{18}) and (\ref{26}) was solved by the
iteration method. The obtained quasiparticle spectrum, for $\sigma(k)$ in a
zeroth-order approximation, is shown in Fig.~3. The spectrum of He-II in the
indicated approximations has been found earlier in work \cite{8}.

We note that the relation $P(\mathbf{k},\mathbf{k}_{1})\sim\frac {\mathbf{k}%
_{1}\mathbf{k}}{k_{1}^{2}}$ at $k\rightarrow0$ and small $k_{1}$ is valid
for $\psi_{\mathbf{k}}$ in a first approximation, which corresponds to the
Feynman-Cohen formula (\ref{2}).

In a second approximation, it is necessary to solve the complete system of
equations (\ref{18})--(\ref{24}). Similarly to work \cite{11}, we solved
these equations numerically. The system of equations (\ref{18}), (\ref{19})
as a whole was solved by the iteration method, while Eq.~(\ref{19}) by the
method of quadratures \cite{29}. In so doing, we used the values of $%
\sigma(k)$ obtained in a first approximation. The error of the numerical
definition of $E(k)$ was about $\pm10~\%$; another error of about $\pm6~\%$
stemmed from measuring $S(k)$ in \cite{24} with an accuracy of $\pm2~\%$.
The obtained spectrum $E(k)$ is shown in Fig.~3.

>From Fig.~3, one can see that if the number of corrections, which are taken
into account, increases, the agreement between the theoretical and
experimental spectra improves, so that for $\Psi_{0}$ and $\psi_{\mathbf{k}}$
determined in a first and a second approximation, respectively, we have a
good agreement between $E(k)$ and the experiment. The \textquotedblleft
shoulder\textquotedblright\ $E(k)\approx17~\mathrm{K}$ in the experimental
spectrum at $k>2.5~\mathrm{\mathring{A}}$ is connected, in our opinion, with
a hybridization of the spectrum that describes a single quasiparticle with a
two-roton level \cite{30}.

Knowing the structure factor, one can restore the interaction potential
between He$^{4}$ atoms by finding $\sigma(k)$ and $f(\mathbf{k}_{1},\mathbf{k%
}_{2})$ from Eqs.~(\ref{6})--(\ref{11}) with known $S(k)$ and substituting
the obtained solutions into the following equation for the Fourier-image $%
\nu(k)$ of the potential \cite{11}:
\begin{align}
& \frac{1}{2}\sigma(k_{1})k_{1}^{2}+\frac{n\nu(k_{1})m}{4\hbar^{2}}%
-\sigma^{2}(k_{1})k_{1}^{2}=  \notag \\
& =\frac{1}{N}\sum\limits_{\mathbf{k}_{2}\neq0,-\mathbf{k}_{1}}f_{s}(\mathbf{%
k}_{1},\mathbf{k}_{2})0.5(k_{2}^{2}+\mathbf{k}_{1}\mathbf{k}_{2}).
\label{27}
\end{align}
The potential
\begin{equation}
U(r)=\frac{1}{(2\pi)^{3}}\int\nu(k)e^{i\mathbf{k}\mathbf{r}}d\mathbf{k},
\label{28}
\end{equation}
where $\nu(k)$ is a solution of Eq.~(\ref{27}), is shown in Fig.~4 for $%
\sigma(k)$ taken in a zeroth-order and a first approximation. The potential $%
U(r)$ in a zeroth-order approximation was obtained in work \cite{23}
earlier. The potential calculated by us approximately agrees with those
obtained in works \cite{5,11,31,32}, but not with Aziz's potential \cite{33}%
, which possesses a very high barrier of repulsion $U(r=0)\sim10^{6}~\mathrm{%
K}$. This discrepancy might be caused by the efficiency of the potential
that describes the interaction between He$^{4}$ atoms at small distances, as
well as by different modeling of such interaction. It is not improbable that
some processes (e.g., the scattering of He$^{4}$ atoms) are better described
by Aziz's potential, while others (in particular, the calculation of $%
\Psi_{0}$, $\psi_{\mathbf{k}}$, and the $E(k)$ spectrum) by a potential with
a much smaller effective barrier $U(0)\sim100~\mathrm{K}$. One can see from
Fig.~4 that the found potential has a minimum at $r_{\mathrm{min}}=3~\mathrm{%
\mathring{A}}$ with the depth $U_{\mathrm{min}}=-7.7~\mathrm{K}$, which
approximately corresponds to the Lennard--Jones experimental
\textquotedblleft well\textquotedblright\ with $r_{\mathrm{min}}=2.97~%
\mathrm{\mathring{A}}$ and $U_{\mathrm{min}}=-10.8$~\textrm{K} \cite{33}.
\begin{figure}[h]
\centerline{\includegraphics[width=85mm]{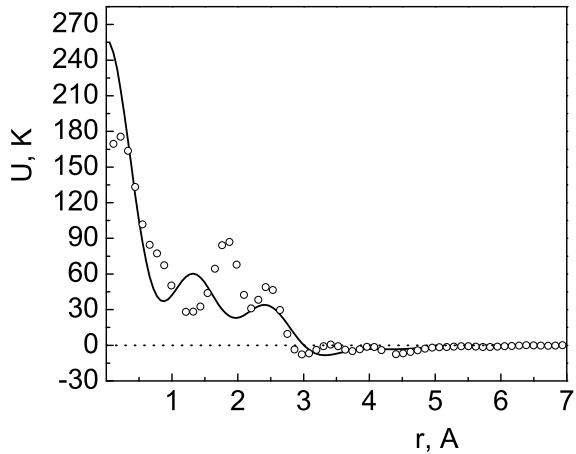}}
\caption{Potential of interaction $U(r)$ between He$^{4}$ atoms,
restored from the structure factor. The solid curve corresponds to
a zeroth-order approximation for $\sigma(k)$, and the circles to a
first one.}
\end{figure}

Fig.~4 also testifies that $U(r)$ for $\sigma(k)$ in a first approximation
differs appreciably from that for $\sigma(k)$ in a zeroth-order one. The
potential changes appreciably even if Eqs.~(\ref{8}) and (\ref{9}) are
rewritten in another but equivalent form. The inaccuracy of the $U(r)$
determination stems from the fact that, according to Eqs.~(\ref{27}) and (%
\ref{28}), the potential $U(r)$ depends strongly on the $\sigma(k)$ values
at $k$'s that are not small, $k=2k_{0}\div4k_{0}$, because the summand $%
\int\sigma(k)k^{4}dk$ makes a contribution to $U(r)$. The values of $%
\sigma(k)$ are small at such $k$'s, but the corrections to $\sigma(k)$ turn
out to be of about $\sigma(k)$ itself. Therefore, $\sigma(k)$ is not
determined exactly at considerable $k$; this circumstance has almost no
influence on the resulting quasiparticle spectrum, but induces a significant
error while finding $U(r)$. Thus, we can only estimate the potential $U(r)$,
but in order to calculate $U(r)$ with a higher accuracy, one must determine
the next approximations for $\sigma(k)$ and measure $S(k)$ more precisely.

\section{Comparison of different He-II models}

Below, we present a short, schematic comparison of various methods which are
applied in order to explain the microstructure of He-II. In so doing, we do
not pretend that our analysis is complete or perfect.

There are plenty of works dealing with the microscopic description of He-II.
Some analysis can be found in reviews \cite{20,34}. The main approaches are
as follows.

\begin{enumerate}
\item[(i)] Semi-phenomenological methods, where certain equations
(like the Gross--Pitayevskii one \cite{35} or that of the model of
a \textquotedblleft continuous medium\textquotedblright\
\cite{36}) are postulated and used as a start point to derive the
quasiparticle spectrum. Several fitting parameters (FPs)
\begin{figure}[h]
\centerline{\includegraphics[width=85mm]{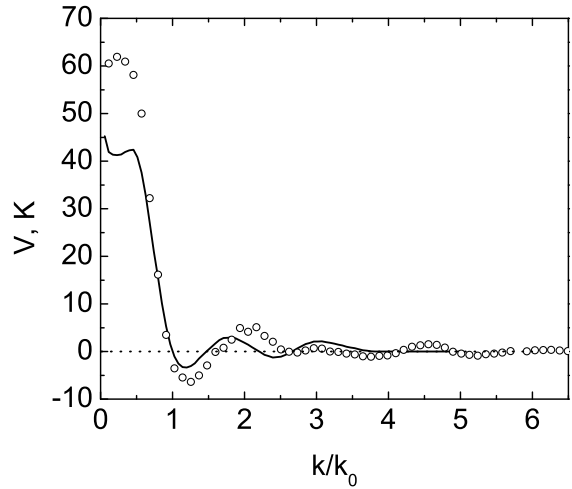}}
\caption{Fourier image $V(k)=n\nu(k)$ for potentials $U(r)$, shown
in Fig.~4,
$k_{0}=2\pi/d$. The notations correspond to the same approximations for $%
\sigma(k)$ as in Fig.~4.}
\end{figure}
are used at that. The main shortcoming of these methods is that it
is not clear how precisely the initial postulates correspond to
the He-II microstructure.

\item[(ii)] Microscopical approaches, which are based on the calculation of $%
\Psi_{0}$ and $\psi_{\mathbf{k}}$.

\begin{enumerate}
\item A \textquotedblleft straightforward\textquotedblright\ solution of the
$N$-particle Schr\"{o}dinger equation in the $\mathbf{r}$- \cite{6} or the $%
\mathbf{k}$-space (the MCV, see Refs.~\cite{8,9,10,11} and this work). Here,
both $\Psi_{0}$ and $\psi_{\mathbf{k}}$ can be determined without
introducing FPs.

\item \textquotedblleft Indirect\textquotedblright\ methods for solving the
Schr\"{o}dinger equation, e.g., the \textquotedblleft correlated basis
function\textquotedblright\ \cite{17} and \textquotedblleft hypernetted
chain\textquotedblright\ \cite{37,38} approaches.

\item Variational methods \cite{1,2,3,4,39}.

\item  In the \textquotedblleft shadow wave
function\textquotedblright\
 (SWF) approach \cite{18,19,20,21}, the attempt is made to partially
``contract'' the whole infinite
           series of correlative corrections to $\ln{\Psi_{0}}$ and
   $\Psi_{\textbf{k}}$ into separate simple ``shadow'' factors.
   This procedure was argued by a certain reasoning, in particular,
   by that taking the delocalization of atoms into account means
   the partial consideration of higher correlations. As was noted in \cite{18},
       such a solution is the first iteration of the
       Schr\"{o}dinger equation represented in the form of a functional
       integral. Drawbacks of the models are as follows: the exact solutions for $\ln{\Psi_{0}}$ and
   $\Psi_{\textbf{k}}$ are infinite series, and it is not clear
   to a which extent the shadow factors will allow one to evaluate this series; moreover, too much FPs are in use.

\item   The numerical Monte-Carlo  (MC) method
\cite{gfmc,qdmc,mor},  which  gives the most exact description of
the ground state,  its energy $E_{0}$,
          the structure factor $S(k)$, and the values of all condensates. But the method  does not allow one to see
     the analytic structure of a solution and its details and  does not yield the curve $E(k)$.

\end{enumerate}

In approaches c, d, several FPs are used.  The main lack of all
models a--d from (II) consists in that
     the exact solutions for $\ln{\Psi_{0}}$ and
   $\psi_{\textbf{k}}$ are infinite correlation series. In practice,
     one succeeds to consider only 2-3 first terms, whereas the omitted corrections are not small.

\item[(iii)] Field-theoretic models.

\begin{enumerate}
\item Studies of the total Hamiltonian $\hat{H}$ in the $\mathbf{k}$-space
\cite{5,14,15,16,40,41}. To a certain extent, this case is rather close to
item (ii,a). The condensates do not appear explicitly in the equations.

\item Studies of the Hamiltonian $\hat{H}$ in the $\mathbf{k}$-space, in the
representation of the operators $\hat{a}_{\mathbf{k}}^{+}$ and $\hat
{a}_{%
\mathbf{k}}$ for quasiparticles \cite{42,43,44}. Here, the condensates
appear explicitly.

\item Solution of the equations similar to the Belyaev--Dyson ones \cite%
{31,32}.
\end{enumerate}

Models b and c involve FPs.
\end{enumerate}

In our opinion, the most perspective may be the field-theoretic
approaches of types (iii,b) and (iii,c), the MC method, or
quantum-mechanical  methods that have not been discovered yet,
which will start from exact microscopic equations, will not use
fitting parameters, and where the expansion in a small parameter
will be carried out. At the same time, approaches (ii) form a
necessary complement to (iii) ones.

\section{Conclusions}

To summarize, in this work, using the method of collective variables, the
spectrum of He-II quasiparticles has been obtained, and the wave functions
of the ground and a first excited state of helium-II have been found
approximately, without introducing any fitting parameter into the model. We
have solved the equations that had been derived from the exact microscopic
equations. A single inaccuracy of the method consisted in breaking the
series for $\Psi_{0}$ and $\psi_{\mathbf{k}}$. The obtained He-II
quasiparticle spectrum agrees well with the experimental one. Therefore, we
believe that the found solution reflects the microstructure of He-II. This
result makes the solution found in work \cite{8} earlier more accurate.

The author is grateful to V.E.~Kireev for discussion of numerical methods
and to E.A. Pashitskii for discussion of the work and useful criticism of
previous results.

\bigskip
[Ukr. J. Phys. -- 2005. -- 50, N 7. -- P. 720 -- 726; Received
11.08.04]

\end{document}